\documentclass[journal,twoside,web]{ieeecolor}
\usepackage{tmi}
\usepackage{cite}
\usepackage{amsmath,amssymb,amsfonts}
\usepackage{algorithmic}
\usepackage{graphicx}
\usepackage{textcomp}
\usepackage{comment}
\usepackage{lineno}
\usepackage{float}
\usepackage{siunitx}
\usepackage[dvipsnames]{xcolor}
\usepackage[utf8]{inputenc}
\usepackage{nccmath}
\def\BibTeX{{\rm B\kern-.05em{\sc i\kern-.025em b}\kern-.08em
    T\kern-.1667em\lower.7ex\hbox{E}\kern-.125emX}}
\begin{document}
\title{Deformable Slice-to-Volume Registration for
Motion Correction of Fetal Body and Placenta MRI}
\author{Alena Uus, Tong Zhang, Laurence H. Jackson, {Thomas A. Roberts}, Mary A. Rutherford, Joseph V. Hajnal and Maria Deprez%
\thanks{Copyright (c) 2020 IEEE. Personal use of this material is permitted. However, permission to use this material for any other purposes must be obtained from the IEEE by sending a request to pubs-permissions@ieee.org. The final version of record is available at http://dx.doi.org/10.1109/TMI.2020.2974844. }
\thanks {This work was supported by the NIH Human Placenta Project grant [1U01HD087202-01], the Wellcome EPSRC Centre for Medical Engineering at King's College London (WT 203148/Z/16/Z), the Wellcome Trust and EPSRC IEH award [102431] for the iFIND project and by the National Institute for Health Research (NIHR) Biomedical Research Centre based at Guy’s and St Thomas’ NHS Foundation Trust and King’s College London.}
\thanks {All authors are with the School of Imaging Sciences \& Biomedical Engineering, King's College London, King’s Health Partners, St. Thomas’ Hospital, London SE1 7EH, United Kingdom (e-mail: alena.uus@kcl.ac.uk).}
}

\maketitle

\begin{abstract}
In in-utero MRI, motion correction for fetal body and placenta poses a particular challenge due to the presence of local non-rigid transformations of organs caused by bending and stretching. The existing slice-to-volume registration (SVR) reconstruction methods are widely employed for motion correction of fetal brain that undergoes only rigid transformation. However, for reconstruction of fetal body and placenta, rigid registration cannot resolve the issue of misregistrations due to deformable motion, resulting in degradation of features in the reconstructed volume. We propose a Deformable SVR (DSVR), a novel approach for non-rigid motion correction of fetal MRI based on a hierarchical deformable SVR scheme to allow high resolution reconstruction of the fetal body and placenta. Additionally, a robust scheme for structure-based rejection of outliers minimises the impact of registration errors. The improved performance of DSVR in comparison to SVR and patch-to-volume registration (PVR) methods is quantitatively demonstrated in simulated experiments and 20 fetal MRI datasets from 28-31 weeks gestational age (GA) range with varying degree of motion corruption. In addition, we present qualitative evaluation of 100 fetal body cases from 20-34 weeks GA range. 
\end{abstract}


\section{Introduction}
\label{sec:intro}

\IEEEPARstart{O}{ver} the past two decades development of fast acquisition sequences along with advanced motion compensation techniques \cite{Malamateniou2013} has gradually allowed incorporation of MRI into clinical practice for imaging of fetal pathologies \cite{Story2015}.

\begin{figure}[!ht]
\centering\includegraphics[width=8.8cm]{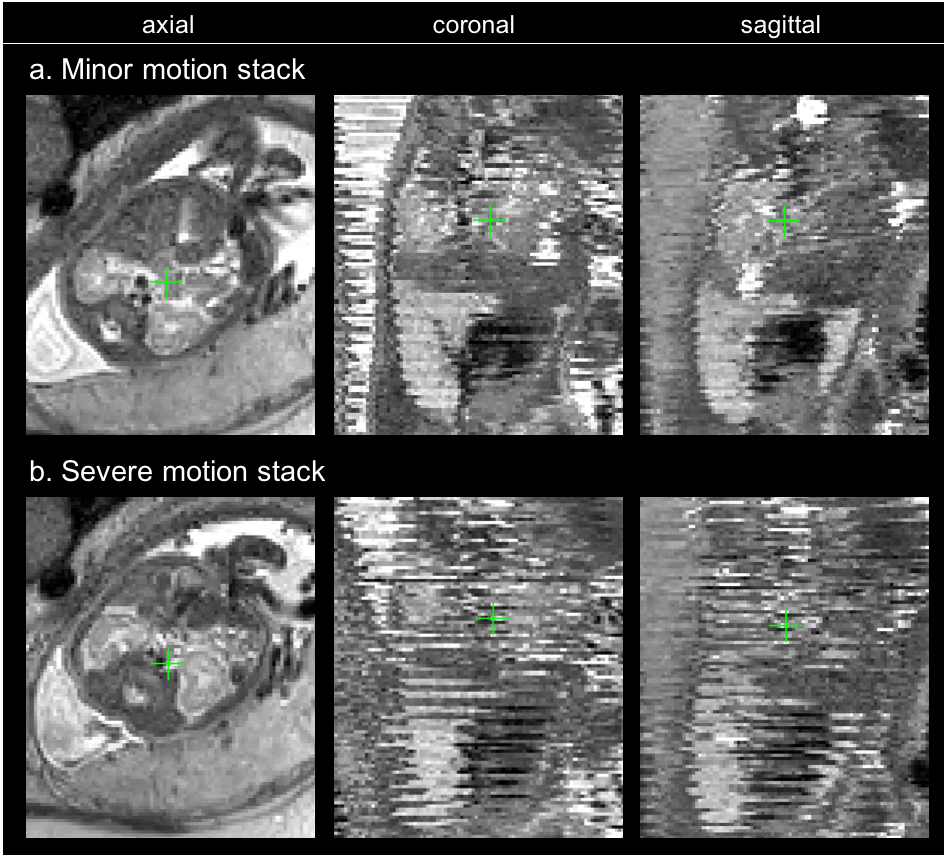}
\caption{ Fetal MRI: examples of stacks with minor (a) and severe (b) motion corruption acquired during the same fetal exam and under the same orientation at different time points. }
\label{fig_1}
\end{figure}

Single shot fast spin echo (ssFSE) sequences allow acquisition of each slice in less than one second, which minimises the impact of fetal motion artefacts on image quality. However, in 3D stacks, inter-slice motion still exists leading to either minor misalignments (Fig.~\ref{fig_1}.a) or complete loss of volumetric information (Fig.~\ref{fig_1}.b). 

\par Slice-to-volume registration in combination with super-resolution (SR) reconstruction is considered to be an efficient motion correction approach since it resolves out-of-plane motion \cite{Gholipour2010, Rousseau2010, Kuklisova-Murgasova2012}. The fact that the volumetric region of interest (ROI) is oversampled at different stack orientations ensures consistency of reconstructed volumes. Recent validation of SVR for fetal brain reconstruction showed strong correlation between 2D and 3D biometry values \cite{Kyriakopoulou2017}. 

\par
\begin{figure}[!h]
\centering\includegraphics[width=8.8cm]{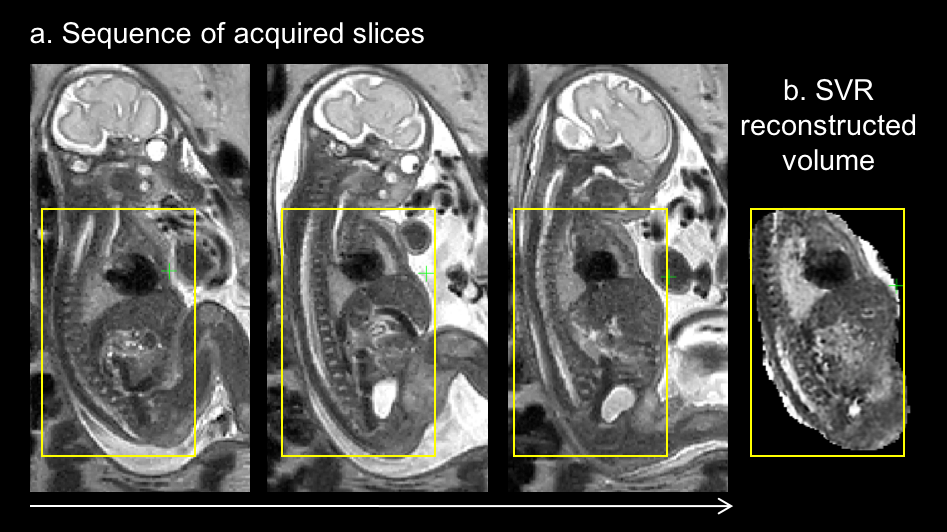}
\caption{Example of a sequence of slices (a) acquired during the change of fetal trunk position and the corresponding SVR reconstructed volume (b). Note that bending deforms the shape of the spine and internal organs in both in-plane and through-plane directions.}
\label{fig_2}
\end{figure}

 Since the SVR is based on rigid registration, its application has primarily focused on the brain as it undergoes only rigid motion. In general, fetal body and placenta are affected by local non-rigid deformations \cite{Nowlan2015}, for example due to bending of fetal abdomen (Fig.~\ref{fig_2}.a) or maternal breathing, and the use of rigid registration leads to deteriorated reconstruction quality. Slices that are inconsistent with the most prevalent body position are either rejected as outliers \cite{Kuklisova-Murgasova2012} or contribute as an error to the reconstructed volume, resulting in  blurring of local features (e.g., spine) and loss of texture information (Fig.~\ref{fig_2}.b).  

\subsection{Related work} 
The original concept of SVR for reconstruction of fetal brain from motion-corrupted MRI stacks was to interleave slice-to-volume registration with scattered data interpolation based on  weighed sum of Gaussian kernels \cite{Rousseau2006} or multilevel B-splines \cite{Jiang2007}. The SVR reconstruction framework was gradually extended with SR reconstruction \cite{Gholipour2010, Rousseau2010}, edge-preserving regularisation \cite{Rousseau2010}, outlier rejection \cite{Gholipour2010, Kuklisova-Murgasova2012}, intensity matching \cite{Kuklisova-Murgasova2012}, total variation regularisation \cite{Tourbier2015}, sinc PSF model and GPU-parallelisation \cite{Kainz2015}. More recent works proposed to employ convolutional neural network (CNN) for fetal brain segmentation from MRI slices \cite{Salehi2018, Ebner2018} and prediction of transformations of slices to canonical atlas space \cite{Hou2018, Salehi2019}. 

\par Though SVR has been developed primarily to reconstruct the fetal brain MRI, it has also been applied to the reconstruction of fetal thorax \cite{Kainz2014, Lloyd2019}  under the assumption of approximately rigid motion, due to the embedding in the rib cage. Given the known cardiac phases of each of the slices, SVR can also be employed for 4D fetal cardiac reconstruction from dynamic MRI \cite{VanAmerom2018}. 
Rigid SVR has also been applied to reconstruct MRI of placenta \cite{TORRENTSBARRENA2019263}, which undergoes deformation due to the maternal breathing. Here the quality of the reconstructed image relies on the robust statistics to exclude the slices which have been deformed compared to the most common shape of the placenta present in the data. In \cite{Alansary2017}, rigid SVR was further extended to patch-to-volume registration (PVR) which addresses the reconstruction of deformable organs by performing motion compensation in piecewise rigid fashion, thus allowing the reconstruction of the entire uterus. 

\par With respect to application of deformable SVR for motion correction, the existing solutions primarily focus on registration of intra-operative slices with a pre-operative planning volume \cite{Tadayyon2011, Xu2015} or multimodal registration (e.g., histology to MRI) \cite{Osechinskiy2010, Rivaz2014}. The majority of monomodal methods are based on rigid SVR for global alignment followed by {Free Form Deformation (FFD) registration \cite{Rueckert1999}} for correction of non-rigid shape changes. Recently, \cite{Ferrante2018} formalised deformable graph-based SVR approach validated on a 3D heart MRI dataset. However, the existing implementation is limited to in-plane deformations only. Model-based SVR methods integrating biomechanical models for physics-based regularisation were proposed in works of \cite{Nir2013, Kuklisova-Murgasova2018}.

\subsection{Contributions}

\par In this paper we present a novel approach for non-rigid motion correction in 3D volumes based on an extension of the rigid SVR reconstruction method \cite{Kuklisova-Murgasova2012} with hierarchical deformable registration scheme and structure-based outlier rejection. The proposed DSVR method allows for both in-plane and out-of-plane correction of local non-rigid deformations of fetal body and placenta resulting in high quality volumetric reconstruction. The structure-based outlier rejection ensures that the residual registration errors are not propagated to the reconstructed volume. 

\par The performance of DSVR is evaluated in simulated experiments, and using 20 fetal MRI datasets from 28-31 weeks GA range with varying degrees of motion corruption, by comparison to the rigid SVR and PVR. Furthermore, we present qualitative evaluation of DSVR reconstruction quality on 100 iFIND\footnote{iFIND Project: http://www.ifindproject.com} cases from 20-34 weeks GA range.

\section{Background}
\label{sec:background}

\par A typical fetal MRI dataset consists of multiple motion corrupted stacks $S=\{S_l\}_{l=1,...,L}$  covering a specific ROI and acquired under different orientations. In the context of motion correction, SVR is used for iterative recovery of a high resolution 'motion-free' isotropic volume $X$ from an array of $K$ low resolution motion-corrupted slices $Y=\{Y_k\}_{k=1,...,K}$. It is formalised as follows \cite{Kuklisova-Murgasova2012}:

\begin{equation} \label{eq1}
Y_k^* = M_k X,\quad y_{jk}^* = s_ke^{-b_{jk}}y_{jk}, 
\end{equation}
where $k$ is a slice index and $j$ is a pixel index within each slice. The matrix $M_k$ describes spatial relationship between the acquired slices $Y_k=\{y_{jk}\}_{j=1,...,N_k}$ and the reconstructed volume $X=\{x_i\}_{i=1,...,N}$, where each row $\{m_{ij}^k\}_{i=1,...,N}$ represents the acquisition PSF for voxel $y_{jk}$, transformed according to the estimated motion parameters and sampled on the grid of the high resolution volume $X$. Intensity corrected slices are denoted by $Y_k^*=\{y_{jk}^*\}_{j=1,...,N_k}$, while $B_k=\{b_{jk}\}_{j=1,...,N_k}$ and $s_k$ are the slice-dependent bias fields and scaling factors, correspondingly. 

\par The algorithm proceeds by interleaving slice-to-volume registration and super-resolution reconstructions for a fixed number of iterations ($q=1,...,Q$). At each SVR iteration $q$, the current estimation of $X$ is registered to slices $\{Y_k\}$ and the transformed PSFs $\{m_{ij}^{k(q)}\}_{q=1,...,Q}$ are updated according to the resulting spatial transformations. Next, as initialisation of the SR reconstruction loop, the weighted Gaussian interpolation \cite{Rousseau2006} is performed for estimation of  {$X^{(0,q)}$. Then, we perform super-resolution reconstruction \cite{Gholipour2010} of the volume $X$ by iterative gradient descent optimisation} based on minimisation of the sum of squared errors $\sum_{jk}e_{jk}^2+\lambda R(X)$, where $\lambda R(X)$ is an edge preserving regularisation term \cite{Kuklisova-Murgasova2012} and $e_{jk}$ are the errors between the original $\{y_{jk}^*\}$ and simulated $\bar{Y_k}=\{\bar{y}_{jk}\}_{j=1,...,N_k}$ slices
\begin{equation} \label{eq2}
 e_{jk}=y_{jk}^*-\bar{y}_{jk},\quad 
 \bar{y}_{jk} = {\sum_{i}}m_{ij}^{k(q)}x_i
\end{equation}
Each SR reconstruction iteration {$(n=1,...,N_{SR}^{(q)})$} includes expectation-maximization (EM) robust statistics scheme for rejection of outliers and estimation of $\{b_{jk}\}$ and $s_k$

\begin{equation} \label{eq3}
x_i^{(n+1,q)}=x_i^{(n,q)}+\alpha{\sum_{kj}} p_{jk} p_k^{slice} m_{ij}^{k(q)}e_{jk}+\alpha\lambda\frac{\partial}{\partial x_i}R(X), 
\end{equation} 
where $p_{jk}$ and $p_k^{slice}$ are posterior probabilities of a voxel or a slice not being an outlier.

\section{Method}
\label{sec:method}

\subsection{Overview of the algorithm}
\label{sec:SUMMARY}
\par
\begin{figure}[!h]
\centering\includegraphics[width=8.8cm]{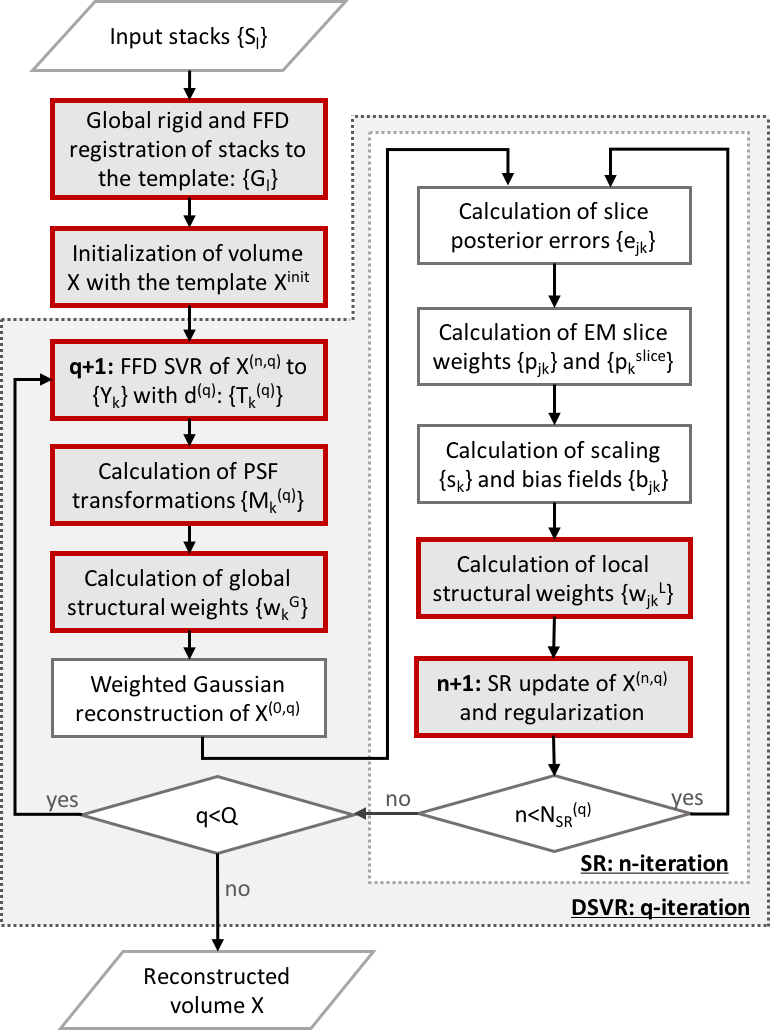}
\caption{Proposed DSVR reconstruction algorithm. The novel elements are highlighted by red outline.}
\label{fig_3}
\end{figure}

 The proposed DSVR method is an extension of the rigid SVR SR reconstruction framework \cite {Kuklisova-Murgasova2012} described in Sec.~\ref{sec:background} and it is summarised in Fig. \ref{fig_3} with the novel modules highlighted by red outline.
\par Following global rigid and FFD registration of all stacks, the template stack is used for initialisation of the registration target for the DSVR loop. At each DSVR iteration (Sec.~\ref{sec:HR}), the current estimation of $X^{(n,q)}$ is registered to each of the slices $\{Y_k\}$. This is followed by the Weighted Gaussian reconstruction and iterative SR reconstruction (Sec.~\ref{sec:SR}) with additional structure-based outlier rejection (Sec.~\ref{sec:OR}) step. The pipeline is performed for a predefined number of DSVR ($Q$) and SR ($N_{SR}$) iterations. The rest of this section describes the novel DSVR modules in detail.

\subsection{Incorporating deformations into SR reconstruction}
\label{sec:SR}

In super-resolution SVR (Sec.~\ref{sec:background}), the forward problem is modelled by applying the transformed PSFs $\{m_{ij}^k\}_{i=1,...,N}$ to high-resolution volume $X$ to simulate (intensity-corrected) voxel $\bar{y}_{jk}$ of the acquired slice $k$ \eqref{eq2}. The underlying PSF in the space of acquired data is a continuous function $f_{jk}(u)=f(u-u_{jk})$ where $u$ is a location in the space of the acquired stack, $u_{jk}$ is the position of the voxel $y_{jk}$ and $f$ is the shape of the acquisition PSF, which we model by a 3D Gaussian with zero mean and principal axis aligned with the axis of coordinate system of the acquired image. The transformation $T_k$ between locations $u$ in space of acquired stack and the anatomical locations $v$ is estimated by registration of the acquired slice $Y_k$ and the volume $X$, defining the transformed PSFs by $m_{ij}^k = f_{jk}(T_k^{-1}(v_i))$, where $v_i$ is {the} location of the voxel $x_i$ in the anatomical space. In case of rigid SVR, the transformed PSFs are re-oriented Gaussians. To correct for deformation of the fetal body and placenta the transformations $T_k(u)$ need to be deformable. Therefore in DSVR the PSFs are deformed using non-rigid transformations $T_k(u)$ and their shapes become non-Gaussian. The high resolution volume $X$ is estimated iteratively using \eqref{eq3} for both rigid and deformable cases.

\subsection{Hierarchical motion correction}
\label{sec:HR}

\par SVR of the fetal brain can be well constrained by acquiring several stacks $\{S_l\}_{l=1,...,L}$ in different orientations and assuming rigid motion of the region of interest, to recover the 'true' shape of the fetal brain. The fetal body and placenta, on the other hand, undergoes continuous deformation in time and DSVR is under-constrained in comparison to SVR. In order to overcome this limitation, a hierarchical scheme for gradual refinement of transformation during slice-to-volume registration is proposed. 

\subsubsection{Volumetric registration} At first, one of the stacks is selected as a template ($S_{template}$) for initialisation of the global registration target. In order to eliminate the impact of global rotations and translations between stacks, the initial step includes 3D-to-3D rigid registration of the input stacks $\{S_l\}$ to a masked ROI in the template stack. The resulting transformations $G_l^{R}$ are then used for initialisation of 3D-to-3D global deformable registration of stacks to $S_{template}$. The template stack cropped with a bounding box of the mask acts as the preliminary initialisation of the reconstructed volume $X^{init}$. 

\par  For the purpose of formalisation of DSVR registration steps, we define a deformable registration operator as $\mathfrak{D}(I_{target}, I_{source}, T_{init}, d)$, where $I_{target}$ and  $I_{source}$ are the source and target images, $d$ represents the resolution of the deformable transformation and $T_{init}$ is the input transformation. Then the global deformable stack registration is expressed as: 

\begin{equation} \label{eq4}
G_l^{D} = \mathfrak{D} (S_l, X^{init}, G_l^{R}, d^{init})
\end{equation}

\par In order to avoid over-fitting to the motion corrupted features of stacks, global transformation with low resolution $d^{init}$ is chosen and all stacks are smoothed using Gaussian blurring. Therefore, the output transformations {$G_l^{D}$} provide estimation of only large range (global) deformations between the trunk positions in the stacks and the template. The trunk mask is then transformed to all stacks and they are cropped to large bounding box ROIs. 

\subsubsection{Slice-to-volume registration} 
\begin{figure}[!h]
\centering\includegraphics[width=8.8cm]{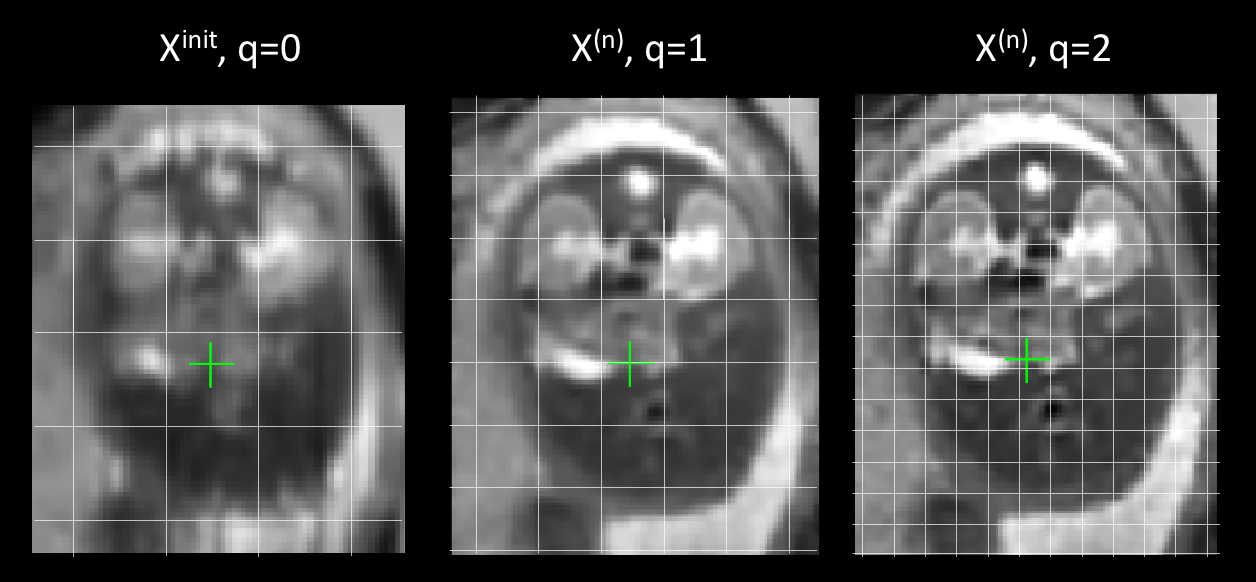}
\caption{An example of refinement of B-spline control point grid ($d^{(q)}$) at each DSVR iteration (sagittal plane with respect to uterus).}
\label{fig_4}
\end{figure}

We refine the motion parameters by iteratively aligning the reconstructed volume to each individual slice. The volume is registered to the slices (i.e., 3D-to-2D) rather than the slices to the volume to ensure that both in-plane and out-of-plane deformable motion is resolved. The first iteration of DSVR is performed by deformable registration of the smoothed template stack $X^{init}$ to each of the slices with low resolution transformations \eqref{eq5}. 
\begin{equation} \label{eq5}
T_{k}^{(0)} = \mathfrak{D}(Y_k, X^{init}, G_{k,l}^{D}, d^{(0)})  
\end{equation}
The B-spline control point spacing $d^{(q)}$ is decreased at every DSVR iteration $q$, thus refining the resolution of the transformation $T_k^{(q)}$. \begin{equation} \label{eq6}
T_{k}^{(q+1)} = \mathfrak{D}(Y_k, X^{(n,q)}, T_{k}^{(q)}, d^{(q+1)})
\end{equation}
This is coupled with decreasing SR regularisation parameter $\lambda$ \eqref{eq3} to prevent overfitting to the residual motion in the early stages and progressively allowing more localised deformations as the features in $X^{(n,q)}$ become better defined. Fig.~\ref{fig_4} illustrates an example of the refinement of {3D} transformation (in our case implemented by a B-spline control point grid) of $T_k^{(q)}$ with respect to DSVR iteration $q$ and reconstructed volume $X^{(n,q)}$ used as a template. It was identified experimentally that rigid registration should not be used during slice-to-volume registration steps since it disrupts the local deformation fields obtained during hierarchical refinement. The spatial relationship coefficients $M_k^{(q)}$ are computed during each iteration $q$ after slice-to-volume registrations were completed, by transforming 3D Gaussian PSFs $f_{jk}$  using $T_k^{(q)}$. 

\subsection{Structure-based outlier rejection}
\label{sec:OR}

\subsubsection{Global structure-based outlier rejection} During each iteration $q$, after the registration and prior to the Gaussian reconstruction step (see Fig.~\ref{fig_3}), misregistered slices are removed  to ensure that global registration errors are not propagated into the initial estimation and further SR reconstruction loop. The quality of registration is assessed as global normalised cross correlation (NCC) between the original slice $Y_k^*$ and the current estimation of the output volume $X^{(n,q)}$ transformed with $T_k^{(q)}$ within the masked slice ROI. The slices with transformations resulting in correlation lower than a threshold value $T^{NCC}$ are excluded. The corresponding slice outlier criteria are computed as: 

\begin{equation} \label{eq7}
    w_{k}^{G} =
    \begin{cases}
      1, & \text{if}\ \frac{1}{N_k} {\sum_{j=1}^{N_k}}\frac{(y_{jk}^*-\mu_{y^*})(x_{jk}^T-\mu_
{x^n})}{\sigma_{y^*}\sigma_{x}} > T^{NCC} \\
      0, & \text{otherwise}, 
    \end{cases}
\end{equation}

where $\{x_{jk}^T=X^{(n,q)}(T_k^{(q)}(u_{jk}))\}_{j=1,...,N_k}$ is the transformed high resolution volume resampled on the grid of the slice $Y_k$. The values $\mu_{y^*}$, $\mu_{x}$, $\sigma_{y^*}$ and $\sigma_x$ are the corresponding intensity means and standard deviation of $\{x_{jk}^T\}$ and $Y_k^*$. \par

\subsubsection{Local structure-based outlier rejection} DSVR of deformable objects can be prone to regional misregistrations and overfitting. Therefore, an additional step for regional outlier rejection is introduced. At each SR iteration $n$ (see Fig.~\ref{fig_3}), the regions of simulated slices with low structural similarity are excluded from contribution to the reconstructed volume. It is based on local structural similarity (SSIM) maps $\{sm_{jk}\}$ between the simulated and original slices: 
\begin{equation} \label{eq8}
    sm_{jk} = \frac{(2\mu_{r^*}\mu_{\bar{r}}+c_1)(2\sigma_{r^*\bar{r}}+c_2)}{(\mu_{{r^*}}^2+\mu_{\bar{r}}^2+c_1)(\sigma_{{r^*}}^2+\sigma_{\bar{r}}^2+c_2)},
\end{equation}
where ${r^*}$ and $\bar{r}$ are the regions in the original $Y_k^*$ and simulated $\bar{Y_k}^{(n)}$ slices centered around voxel $j$ with circular window and $\mu_{r^*}$, $\mu_{\bar{r}}$, $\sigma_{r^*}^2$ and $\sigma_{\bar{r}}^2$ are the corresponding average and variance intensity values and $\sigma_{r^*\bar{r}}$ is the covariance of ${r^*}$ and $\bar{r}$. As defined in \cite{Wang2004}, values $c_1$ and $c_2$ used in order to balance the division with weak denominator are computed as $(k_1L)^2$ and $(k_2L)^2$, where $L$ is the dynamic range of intensities in $r^*$ and $k_1$ and $k_2$ are equal to 0.001 and 0.003, correspondingly.
The outlier criteria of slice voxels with similarity $<T^{SSIM}$ are set to zero: 

\begin{equation} \label{eq9}
    w_{jk}^L =
    \begin{cases}
      1, & \text{if}\ sm_{jk} > T^{SSIM}\\
      0, & \text{otherwise}
    \end{cases}
\end{equation}

\subsubsection{Incorporating structure-based outlier rejection into SR reconstruction} 
Outlier rejection is incorporated into SR reconstruction \eqref{eq3} by scaling the contribution of the reconstruction error $e_{jk}$ by a weight $w_{jk}$:
\begin{equation} \label{eq12}
x_i^{(n+1,q)}=x_i^{(n,q)}+\alpha{\sum_{kj}} w_{jk}  m_{ij}^{k(q)}e_{jk}+\alpha\lambda\frac{\partial}{\partial x_i}R(X)
\end{equation}
In our previous work \cite{Kuklisova-Murgasova2012} this weight was set based on EM robust statistics on voxel and slice intensities to $p_{jk} p_k^{slice}$. The proposed total structural outlier criteria for a voxel defining its contribution to the reconstructed volume is computed as:
\begin{equation} \label{eq11}
w_{jk}^S=w_k^G w_{jk}^L
\end{equation}
In Section \ref{sec:par_study} we will show that the combination of the two schemes by setting $w_{jk}=p_{jk} p_k^{slice}w_{jk}^S$ outperforms the individual outlier rejection schemes.

\section{Implementation}

\subsection{Input data requirements}
\label{sec:INPUT}
\par The input dataset includes stacks of different orientations, an approximate mask covering the ROI (i.e., fetal body or placenta) and a selected template, which can be either one of the stacks or a scout scan. The global structure of the target object needs to be preserved in the template stacks, as it is used for initialisation of the registration target. Minor to average degree of motion corruption of the template is acceptable and is resolved by Gaussian blurring. The severity of motion is visually assessed with respect to the degree of loss of volumetric structural information (see Fig.~\ref{fig_1}). 

\par Rigid SVR requires masking of the ROI in the template stack in order to eliminate the impact of the independent motion of the mother and the fetus. On the other hand, FFD registration does not require precise masking and in our experience preforms better for larger ROIs. 

\par Variable orientations of input stacks help to prevent overfitting to a particular motion-corrupted stack. The minimal requirement for SR reconstruction of an isotropic volume from multiple stacks is two sufficiently different orientations. It was identified experimentally that using 5 to 8 stacks is sufficient for good quality reconstruction depending on the amount of motion corruption, resolution and SNR level of the original volumes. Similarly to the capture range limitation of SVR \cite{Kainz2015}, gradient-descent based deformable registration methods are not capable of resolving motion involving large degree rotations or excessive bending, which should be taken into account with respect to selection of input stacks.

\subsection{Deformable registration}
\label{sec:DEFORMABLE}
\par The B-spline FFD registration \cite{Rueckert1999} with NMI similarity measure was chosen for both deformable SVR and global registration steps due to the lower computational requirements compared to the diffeomorphic registration such as FFD parameterised by stationary velocity (SV) fields \cite{Schuh2014}. Although SV FFD ensures invertibility of transformations, it did not lead to an indicative improvement of reconstruction results while significantly increasing processing time, which made it not feasible for our application.  

\par Typically, the reconstruction pipeline requires 3 SVR iterations $(Q=3)$ each of which is followed by 10 to 30 SR ($N_{SR}^{(q)}$) iterations with gradually refined regularisation parameters. The resolution of B-spline FFD transformation is controlled by changing the resolution of the B-spline control point (CP) spacing $d$. { We choose resolution scheme with B-spline control point spacings $d^{(0)}=d^{init}, d^{(1)}=2/3\cdot d^{init}, d^{(2)}=1/3\cdot d^{init}$.} As we show in Sec.~\ref{sec:res}, it was identified experimentally that $15mm\rightarrow10mm\rightarrow5mm$ CP refinement ($d^{init}=15 mm$) produces the optimal reconstruction quality for our fetal test cases and $0.85mm$ output resolution. The corresponding optimal regularisation $\lambda^{(q)}$ values were chosen as $0.1\rightarrow0.05\rightarrow0.02$ similarly to the SVR settings in \cite{Kuklisova-Murgasova2012}.

\subsection{Structure based outlier rejection parameters }
\label{sec:OR-PAR}
\par Analysis of the choice of the structural similarity thresholds showed that the optimal values corresponding to adequate registration quality are $T^{NCC} = 0.75$ for global and $T^{SSIM} = 0.6$ for local regions. Using lower values might lead to inclusion of regions that were erroneously overfitted. 

\par The 20 mm diameter for SSIM kernel was experimentally identified as optimal for the feature sizes in 28-31 weeks GA range subjects, e.g., the transverse diameter of fetal kidneys for this GA range varies within $15 - 25$mm \cite{VanVuuren2012}. 

\subsection{Software packages and hardware requirements}
\label{sec:SOFTWARE}

\par DSVR framework was implemented based on MIRTK\footnote{MIRTK: https://github.com/BioMedIA/MIRTK } library with multi-CPU parallelisation of registration and reconstruction steps. The structure and functionality of the core reconstruction steps follow the  IRTK-based\footnote{IRTK: https://github.com/BioMedIA/IRTK } implementation of the original SVR reconstruction method \cite{Kuklisova-Murgasova2012}. {The code is available online as a part of SVRTK\footnote{SVRTK: https://github.com/SVRTK/SVRTK } package.}

\par The major advantage of MIRTK registration library is the use of conjugate gradient descent optimisation \cite{Modat2009} that significantly increases computational efficiency of FFD registration that constitutes the most time-consuming part of DSVR pipeline. Depending on the ROI size (related to GA of the subjects), number of stacks, output resolution and the system configuration, the reconstruction time can typically vary between 15 to 60 minutes.

\section{Experiments and Results}

\par We evaluate DSVR based on the comparison to the rigid SVR method \cite{Kuklisova-Murgasova2012} that was recently reported to produce the best results for placenta reconstruction \cite{TORRENTSBARRENA2019263} {and its GPU version \cite{Kainz2015} was employed for 3D fetal cardiac reconstruction in \cite{Lloyd2019}.} Furthermore, DSVR is compared to the recently introduced PVR method designed for motion correction in large FoV regions \cite{Alansary2017}. 
The reconstruction quality is evaluated using intensity and structural similarity metrics.

\subsection{Fetal MRI data}

\par The fetal MRI data used for evaluation contains 20 iFIND T2-weighted datasets of fetuses from 28-31 weeks GA range. This particular GA range was selected due to the lower amplitude of movement and higher prevalence of bending and stretching \cite{Nowlan2015}, since this study focuses on correction of local non-rigid deformations of organs rather than global change of fetal body position. 

\par The iFIND acquisitions were performed on a 1.5T MRI using ssFSE sequence with TR = 15000 ms, TE = 80 ms, voxel size = 1.25 x 1.25 x 2.5 mm, slice thickness 2.5 mm and slice spacing 1.25 mm. The stacks were acquired under different orientations, with 100-160 slices per stack, depending on GA and orientation. Each of the datasets contains 6 stacks with minimum 3 different orientations without major SNR loss.

\par The datasets were divided into 2 groups with respect to severity of motion. The minor motion group contains 10 cases that include stacks only with minimal loss of structural information. In 10 cases from the severe motion group, the majority of stacks have severe misalignment of slices. The severity of motion corruption was visually estimated by an operator with respect to the consistency of volumetric information in all three planes (Fig.~\ref{fig_1}) as well as the changes of the global fetal body position between the stacks. Template stack selection was performed manually based on the degree of motion corruption  and the position of the fetal body similar to the other stacks.  

\begin{table}[!h]
\renewcommand{\arraystretch}{1.1}
\centering

\caption{Stack motion corruption assessment: sequential slice NCC. }
\begin{tabular}{|c|c|}
\hline
 {\textbf{Minor motion datasets}} & { \textbf{Severe motion datasets} }\\ 
\hline
 { 0.688 $\pm$ 0.071 }   & { 0.475 $\pm$ 0.130 }  \\ \hline
\end{tabular}
\label{tab_1}

\end{table}

\par The amount of motion in each stack was also assessed as the average NCC between sequential slices for a masked ROI. Tab.~\ref{tab_1} demonstrates that the severe motion datasets have lower average NCC range than the minor motion datasets.

\subsection{Simulated experiment}

\par In order to assess the general capability of DSVR to recover consistent volumetric information and local anatomy features, we perform a phantom experiment with simulated non-rigid motion. At first, a high quality volume reconstructed from a minimal motion dataset is selected as a reference. Next, five sets of slice transformations (incorporating both local non-rigid and global rigid and non-rigid components) extracted from other existing reconstruction cases are used to generate motion-corrupted stacks from the reference volume. 

\par In this experiment, each of the five simulated datasets contains six generated stacks that have different orientations and a mask defining trunk ROI in the template stack. The default DSVR reconstruction pipeline is executed for all datasets. In addition, rigid SVR \cite{Kuklisova-Murgasova2012} and PVR \cite{Alansary2017} reconstructions are performed for comparison to the state-of-the-art methods. 

\begin{figure}[!h]
\centering\includegraphics[width=8.8cm]{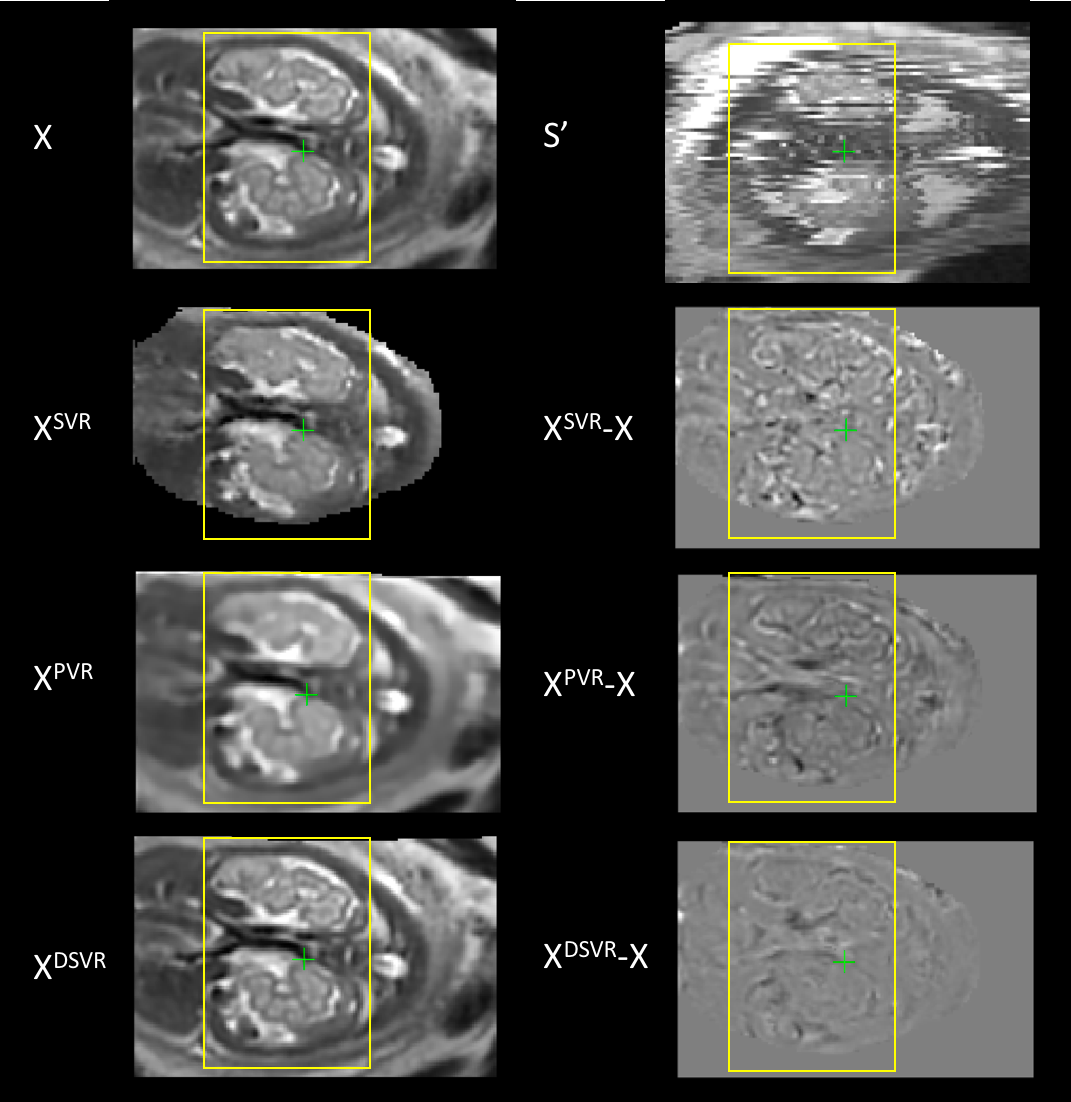} 
\caption{Simulated experiment: original reference volume ($X$), one of the generated motion-corrupted stacks ($S'$), SVR ($X^{SVR}$), PVR ($X^{PVR}$) and DSVR ($X^{DSVR}$) reconstruction results and their difference with the reference (coronal plane).}
\label{fig_5}
\end{figure}

\par Fig.~\ref{fig_5} shows coronal plane view of the original reference volume ($X$), one of the stacks with simulated deformable motion ($S'$), DSVR reconstructed volume ($X^{DSVR}$) along with its difference with the reference ($X^{DSVR}-X$) and SVR ($X^{SVR}$) and PVR ($X^{PVR}$) results. Prior to the analysis of the results, in order to avoid possible impact of the global change of the body position, the reconstructed volumes were aligned to the reference using FFD registration with large CP spacing (15 mm). All three methods successfully reconstructed the major anatomy structures including topology of kidneys and spine. However, in SVR and PVR outputs, misregistrations due to non-rigid deformations led to blurring of texture of local features and higher errors.

\begin{table}[!h]
\renewcommand{\arraystretch}{1.1}
\centering
\caption{Simulated experiment: SVR, PVR and DSVR vs. reference.}  
\begin{tabular}{|c|c|c|c|}
\hline
 \textbf{Method} & \textbf{NRMSE} & \textbf{PSNR} & \textbf{NCC} \\ 
\hline
SVR  & 0.119 $\pm$ 0.002   & 28.916 $\pm$ 0.479  & 0.938 $\pm$ 0.003 \\ \hline
{ PVR } & { 0.182 $\pm$ 0.009 }  & {  25.602 $\pm$ 0.383 } & { 0.882 $\pm$ 0.009 } \\ \hline
DSVR & 0.078 $\pm$ 0.008   & 32.560 $\pm$ 0.859  & 0.973 $\pm$ 0.006 \\ \hline
\end{tabular}
\label{tab_2}
\end{table}

\par The corresponding quantitative comparison of the motion-free reference volume and five DSVR, SVR and PVR reconstructions in terms of normalised root mean square error (NRMSE), peak signal-to-noise ratio (PSNR) and NCC computed for the same masked ROI covering fetal trunk is given in Tab.~\ref{tab_2}. All results are statistically significant with $p < 0.001$. There is a strong correlation between the original and DSVR volumes. The worse results for the SVR outputs indicate that the impact of non-rigid deformations on texture cannot be resolved by rigid registration even with rejection of outliers. The reconstructed PVR volumes also have lower similarity to the reference volumes.

\begin{table}[!h]
\renewcommand{\arraystretch}{1.1}
\centering
\caption{Simulated experiment: SVR vs. DSVR TRE [mm].}  
\begin{tabular}{|c|c|}
\hline
 \textbf{   {     SVR TRE [mm]   }    } & \textbf{  {      DSVR TRE [mm]    }   } \\ 
\hline
{  2.279 $\pm$ 0.509 }  & {  0.797 $\pm$ 0.158 } \\ \hline
\end{tabular}
\label{tab_3}
\end{table}

\par We also calculated target registration error (TRE) to evaluate average displacement error of the estimated non-rigid transformations compared to the the original ones used for generation of motion-corrupted stacks. Comparison of TRE for DSVR and SVR reconstruction is presented in Tab.~\ref{tab_3}. DSVR TRE values are significantly lower in comparison to SVR and statistically significant with $p < 0.001$.

\subsection{Reconstruction of fetal body}
\label{sec:res}

\par In fetal body reconstruction, due to the absence of the ground truth as well as the constantly changing shape of the trunk organs, assessment of the quality of reconstructed volumes is challenging. In \cite{Kuklisova-Murgasova2012}, leave-one-out analysis was proposed for evaluation of SVR results. It is based on the comparison of the original $\{Y_k^*\}$ to simulated $\{\bar{Y_k}\}$ slices for a stack that was registered in SVR step but excluded from SR reconstruction thus not contributing to the output volume.

\par Tab.~\ref{tab_4} presents quantitative comparison of SVR, PVR, DSVR and DSVR with structural outlier rejection (DSVR+S) results for the same masked ROI of the excluded stack for 20 datasets. The values for PVR results are given primarily for a reference, since, due to the differences in implementation, comparison is performed for simulated and original patches with overlapping regions rather than for slices. Furthermore, PVR employs different SR reconstruction pipeline and does not provide an option for stack exclusion.

\begin{table}[!h]
\renewcommand{\arraystretch}{1.1}
\centering
\caption{Fetal body reconstruction. Leave-one-out analysis:   SVR, PVR, DSVR and DSVR+S.}  
\begin{tabular}{|c|c|c|c|}
\hline
\textbf{Method} & \textbf{NRMSE} & \textbf{PSNR} & \textbf{NCC} \\ 
\hline
\multicolumn{4}{|c|}{Minor motion group (10 datasets):}                         \\ \hline
SVR     & 0.229 $\pm$ 0.023   & 24.328 $\pm$ 0.920      & 0.780 $\pm$ 0.084    	\\ \hline
PVR*    & 0.264 $\pm$ 0.050   & 22.438 $\pm$ 1.371      & 0.712 $\pm$ 0.120 	\\ \hline
DSVR    & 0.177 $\pm$ 0.021   & 26.764 $\pm$ 1.139      & 0.863 $\pm$ 0.038 	\\ \hline
DSVR+S  & 0.174 $\pm$ 0.025   & 26.764 $\pm$ 1.249      & 0.867 $\pm$ 0.043 	\\ \hline
\multicolumn{4}{|c|}{Severe motion group (10 datasets):}                        \\ \hline
SVR     & 0.275 $\pm$ 0.025   & 23.219 $\pm$ 0.861      & 0.646 $\pm$ 0.068    	\\ \hline
PVR*    & 0.280 $\pm$ 0.046   & 21.873 $\pm$ 1.436      & 0.633 $\pm$ 0.107 	\\ \hline
DSVR    & 0.222 $\pm$ 0.032   & 25.422 $\pm$ 0.619      & 0.831 $\pm$ 0.043 	\\ \hline
DSVR+S  & 0.214 $\pm$ 0.033   & 25.746 $\pm$ 0.597      & 0.844 $\pm$ 0.040 	\\ \hline
\multicolumn{4}{|c|}{(*) PVR comparison was performed on the patch level.}  \\ \hline
\end{tabular}
\label{tab_4}
\end{table}

\par The results for both minor and severe motion datasets show that DSVR surpasses SVR and PVR for both intensity and structural characteristics. Additional structural outlier rejection (DSVR+S) produces a significant improvement only for the severe motion datasets. This is expected since minor motion assumes high NCC values of registration output and DSVR+S should produce only minimal impact. All results apart for comparison of DSVR and DSVR+S for minor motion cases are statistically significant with $p < 0.005$. 

\begin{figure}[!h]
\centering\includegraphics[width=8.8cm]{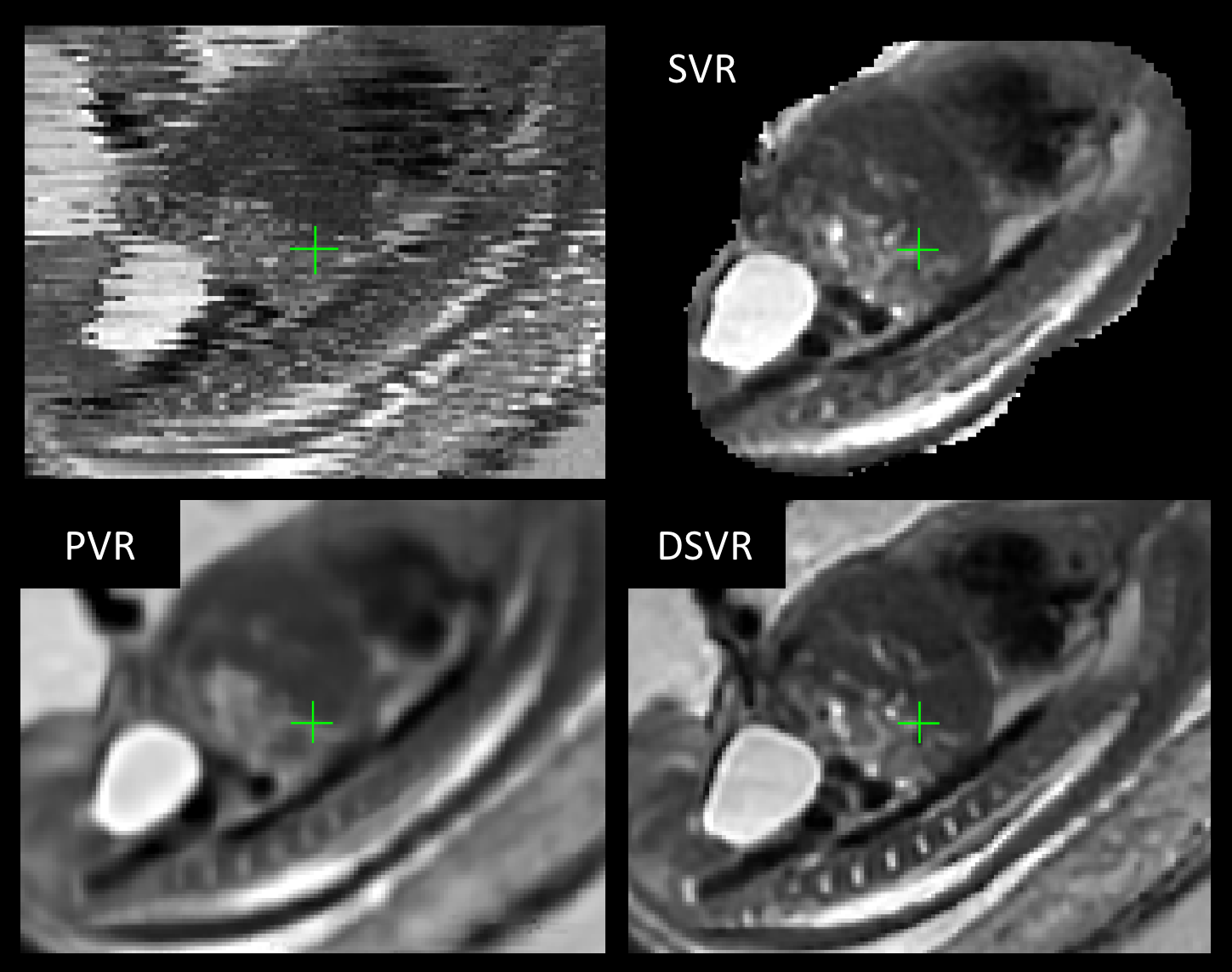}
\caption{Example of motion correction for a minor motion dataset: motion corrupted stack, SVR, PVR and DSVR reconstructions (sagittal plane).}
\label{fig_6}
\end{figure}

For one of the minor motion cases shown in Fig.~\ref{fig_6} (sagittal plane view) when the trunk positions in all stacks are approximately aligned and there are no severe non-rigid deformations, SVR successfully reconstructs the global trunk topology. However, due to bending motion, there is a noticeable loss of structure in the spine region as well as the general degradation of texture. PVR allows reconstruction of the large ROI and partially resolves these artefacts improving definition of the spine. However, the introduced smoothing lowers image quality in terms of interpretation and resolution of small features. On the other hand, DSVR results are characterised by high definition of the local anatomy structures. It also has to be noted that there is a noticeable change in the position of the trunk between different reconstruction methods caused by the different approaches for initialisation of the registration target ($X^{init}$). SVR and PVR use the average of all stacks after global rigid stack registration and DSVR uses the selected template stack. Therefore, the SVR and PVR solutions converge to an intermediate averaged state, whereas DSVR converges to the trunk shape in the template stack.

\begin{figure}[!h]
\centering\includegraphics[width=8.8cm]{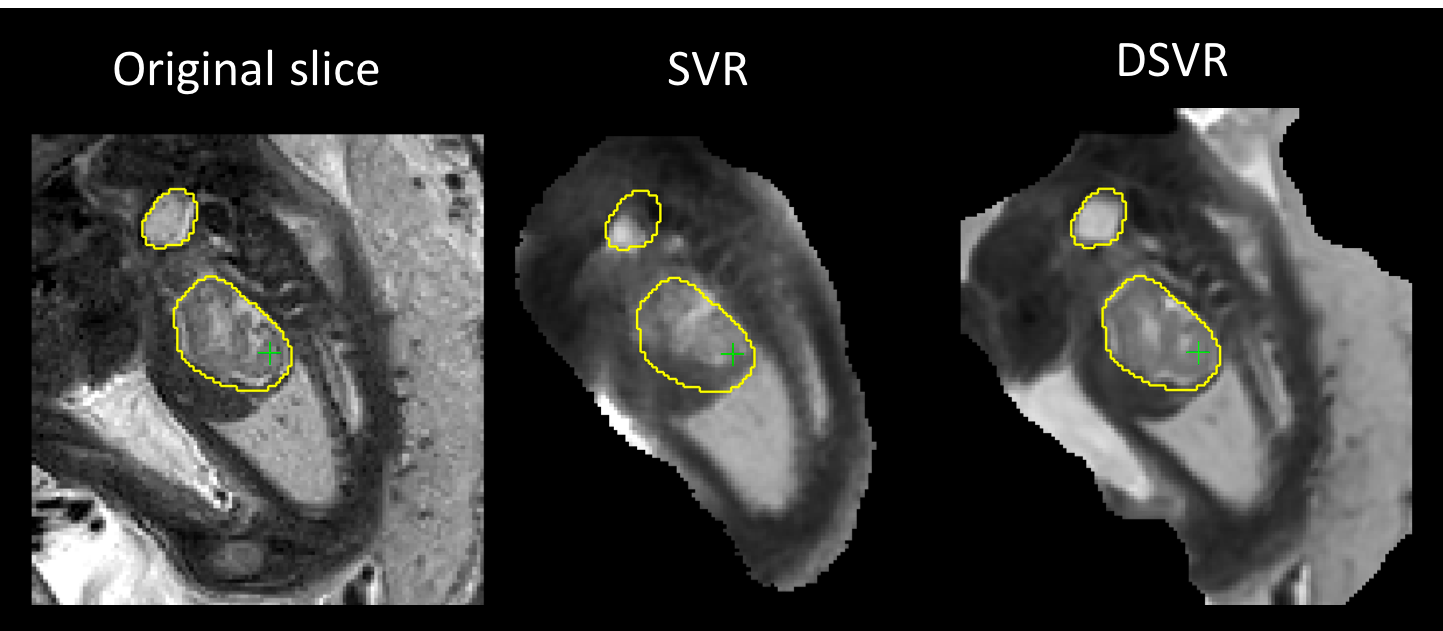}
\caption{Comparison of SVR and DSVR in presence of non-rigid motion: original acquired slice ($Y_k$) vs. slices simulated from SVR and DSVR ($\bar{Y_k}$). The yellow isolines delineate the structure in the original slice, and show misalignment with SVR reconstruction due to limitation of rigid motion correction. The problem is resolved by non-rigid motion correction in DSVR. }
\label{fig_7}
\end{figure} 

\par A typical example of failed rigid SVR due to non-rigid motion is given in Fig.~\ref{fig_7} where one of the original slices $Y_k$ is compared to the corresponding simulated slices $\bar{Y_k}$ from SVR and DSVR reconstructions. The kidney and bladder regions are segmented in order to assess the registration accuracy. In this case, SVR could not correct the impact of spine bending thus converging to an average position with displaced kidney and resulting in large errors $\{e_{jk}\}$. On the other hand, FFD registration improves the mapping ($T_k$) between $Y_k$ and {$X^{(n,q)}$}. 
The deformation of ROI boundaries in DSVR output indicates the high degree of non-rigid deformation. 

\par
\begin{figure}[!h]
\centering\includegraphics[width=8.8cm]{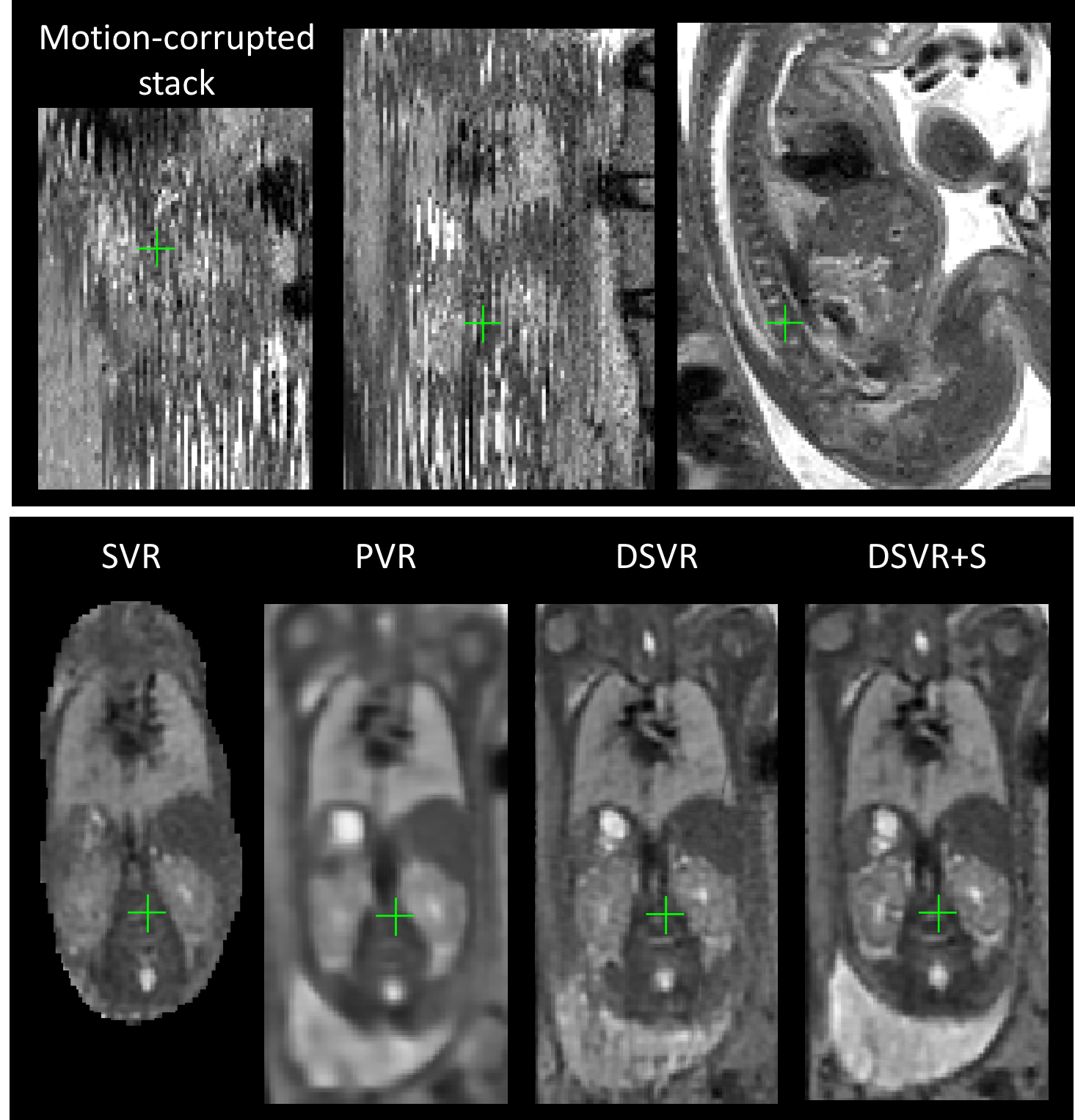}
\caption{Example of motion correction for a severe motion dataset: motion corrupted stack, SVR, PVR and DSVR reconstructions (coronal plane).}
\label{fig_8}
\end{figure}

 For one of the severe motion dataset results shown in Fig.~\ref{fig_8} {(coronal plane view)}, large slice misregistration errors lead to a severe degradation of local features in SVR. PVR resolves this producing a clear trunk structure, however, similarly to the previous example, the smoothed texture lowers the quality of definition of abdominal organs. Although there is an improvement in DSVR vs. SVR output, a significant amount of artefacts due to misregistrations still remains. As mentioned in Sec.~\ref{sec:method}, the employed gradient-descent FFD method is not capable of resolving large bending and rotations therefore leading to misregistrations. 

Structural outlier removal (DSVR+S) improves the output by minimising the contribution of registration errors to reconstruction. It also increases the proportion of rejected slices, which, for severe motion datasets, can vary between 20 - 50 $\%$ of the total slice number. In comparison, the original EM robust statistics \cite{Kuklisova-Murgasova2012} results in only 10 - 15 $\%$ slices being rejected, which seems to be insufficient in major motion cases.

\subsection{ {Parametric study}}
\label{sec:par_study}
\par Regularisation parameters that control the smoothness of the transformations in DSVR have significant impact on the quality of reconstruction. Transformations with large CP spacings allow for correction of the global body shape, but are not efficient for recovery of local features. On the other had, too small CP spacing will lead to overfitting and unrealistic deformations. We experimentally determined that the optimal multi-resolution scheme for fetal body dimensions is $15mm\rightarrow10mm\rightarrow5mm$ for 3 iterations. 

\begin{figure}[!h]
\centering\includegraphics[width=8.8cm]{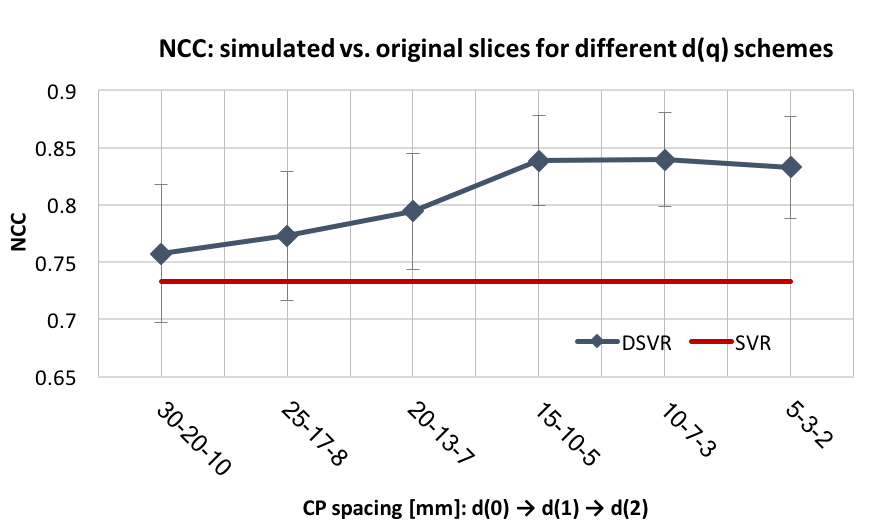}
\caption{ CP spacing analysis: NCC between the original ($Y_k^*$) vs. simulated ($\bar{Y_k}$) slices.}
\label{fig_9}
\end{figure}

\par We evaluated the regularisation schemes with {the CP spacings $d^{(0)}$, $d^{(1)}=2/3 \cdot d^{(0)}$ and $d^{(2)}=1/3 \cdot d^{(0)}$ for the subsequent DSVR iterations.} The $d^{(0)}$ value was varied from $30mm$ to $5mm$ for five severe motion datasets. We calculated NCC between simulated $\{\bar{Y_k}\}$ and original $\{Y_k^*\}$ slices for the masked fetal body ROI in all stacks. Fig.~\ref{fig_9} shows the average NCC values over the five subjects for different initial transformation resolutions $d^{(0)}$. The average SVR output value is provided for the reference. We can observe that $d^{(0)}$=15mm results in optimal performance and further refinement does not improve the results due to overfitting to the motion artifacts. 

\begin{table}[!h]
\renewcommand{\arraystretch}{1.1}
\centering
\caption{Outlier rejection scheme assessment: NCC between ($Y_k$) and ($\bar{Y_k}$) and proportion of excluded slices.}  
\begin{tabular}{|c|c|c|}
\hline
     { \textbf{Method} } & {  \textbf{NCC} } & { \textbf{\% of excluded slices} } \\ 
\hline
{ EM }                & {  0.828 $\pm$ 0.028  } & {  13.41 $\pm$ 2.88 $\%$ } \\ \hline
{ EM + G-STR }         & {  0.839 $\pm$ 0.037  } & { 38.51 $\pm$ 15.28 $\%$ } \\ \hline
{ EM + L-STR }       & { 0.834 $\pm$ 0.031  } & { 18.85 $\pm$ 5.17 $\%$ } \\ \hline
{ EM + L-STR + G-STR } & { 0.841 $\pm$ 0.036  } & { 35.59 $\pm$ 12.29 $\%$ } \\ \hline
{ L-STR + G-STR } & { 0.837 $\pm$ 0.039  } & { 30.19 $\pm$ 11.06 $\%$ } \\ \hline
\end{tabular}
\label{tab_5}
\end{table}

\par In addition, we performed quantitative assessment of the impact of the structural outlier rejection steps on the reconstruction quality and number of excluded slices. Tab.~\ref{tab_5} presents the results for 5 severe motion datasets with the default processing pipeline and different combinations of outlier rejection methods: EM-based method proposed in \cite{Kuklisova-Murgasova2012}, global structural slice-level rejection based on global NCC (G-STR), and local region rejection method based on SSIM maps (L-STR). We report NCC between original $Y_k$ and simulated slices $\bar{Y_k}$ of the excluded stack within the masked fetal body ROI, and the average proportion of the excluded slices. It can be observed that combination of all three outlier rejection steps results in best performance in severe motion cases.

\subsection{Reconstruction of placenta}
\label{sec:res_pl}

\par
\begin{figure}[!h]
\centering\includegraphics[width=8.8cm]{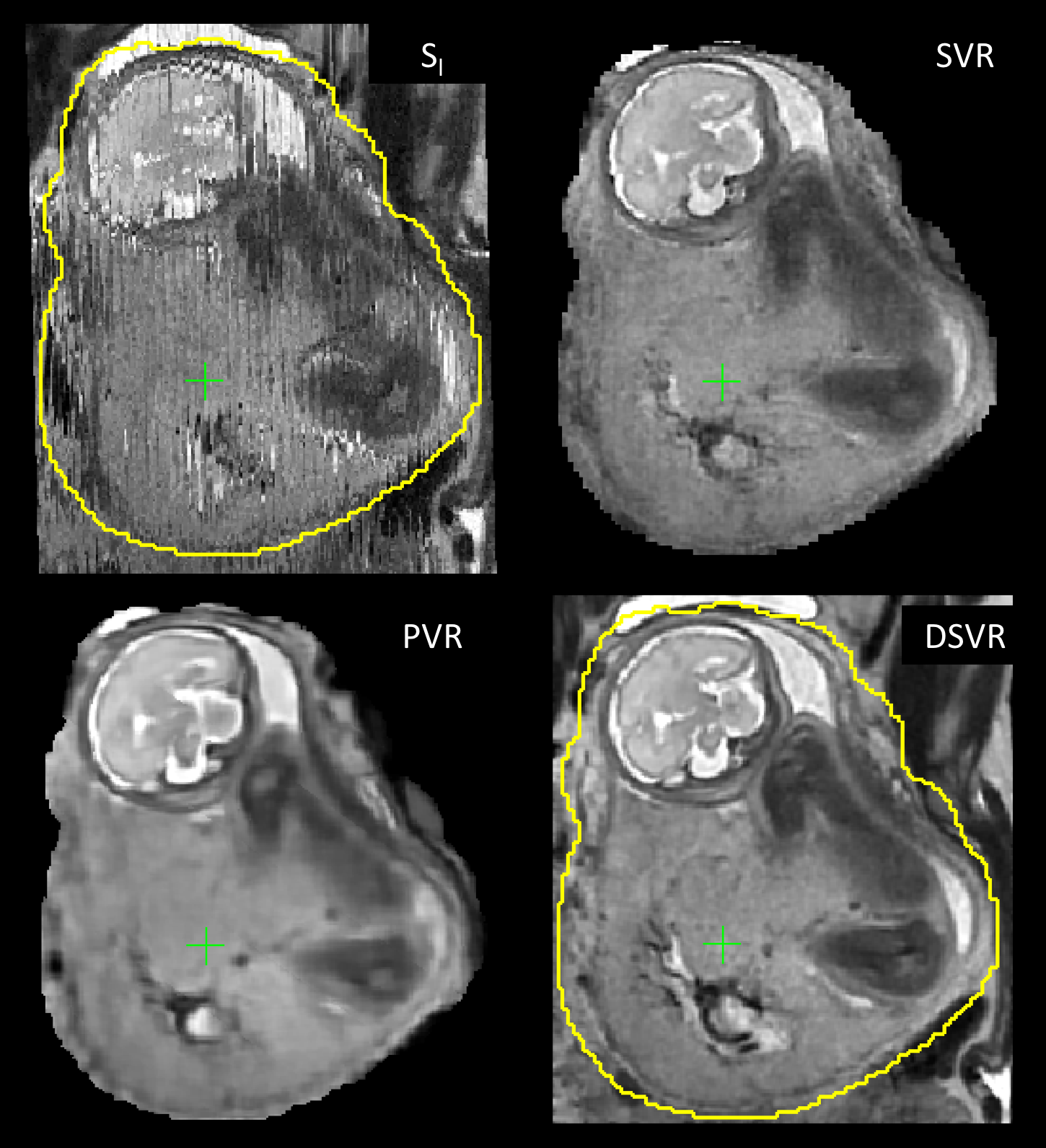}
\caption{Example of motion correction for placenta: motion-corrupted stack, SVR, PVR and DSVR+S reconstructed volumes (coronal plane).}
\label{fig_10}
\end{figure}

 Furthermore, the performance of SVR, PVR and DSVR for reconstruction of placenta was compared for 10 iFIND fetal MRI cases. Each of the datasets contains 5 orthogonal stacks covering the entire uterus and reconstructions are performed for the masked uterus ROI similarly to \cite{TORRENTSBARRENA2019263}. The corresponding results of the leave-one-out analysis are presented in Tab.~\ref{tab_6} in terms of NCC between the original  $Y_k$ and simulated $\bar{Y_k}$ slices of the same masked ROI of an excluded stack. The proposed DSVR method outperforms both other methods. All results are statistically significant with $p < 0.005$. Fig.~\ref{fig_10} shows an example of motion-corrected outputs of SVR, PVR and DSVR of placenta. Placenta is primarily affected by respiratory motion, and is subject to stretching and bending. As a result, DSVR provides better reconstruction quality compared to the alternative methods.

\begin{table}[!h]
\renewcommand{\arraystretch}{1.1}
\centering
\caption{Placenta reconstruction. Leave-one-out analysis:   SVR, PVR and DSVR (NCC).}  
\begin{tabular}{|c|c|c|}
\hline
 \textbf{ { SVR } } & \textbf{ { PVR* } } & \textbf{ { DSVR+S } } \\ 
\hline
{  0.639 $\pm$ 0.085 }  & {  0.726 $\pm$ 0.045 }  & {  0.792 $\pm$ 0.072 } \\ \hline
\multicolumn{3}{|c|}{(*) PVR comparison was performed on the patch level.}  \\ \hline
\end{tabular}
\label{tab_6}
\end{table}

\subsection{Qualitative analysis}
\label{sec:res_pl}

\par \par
\begin{figure}[!h]
\centering\includegraphics[width=8.8cm]{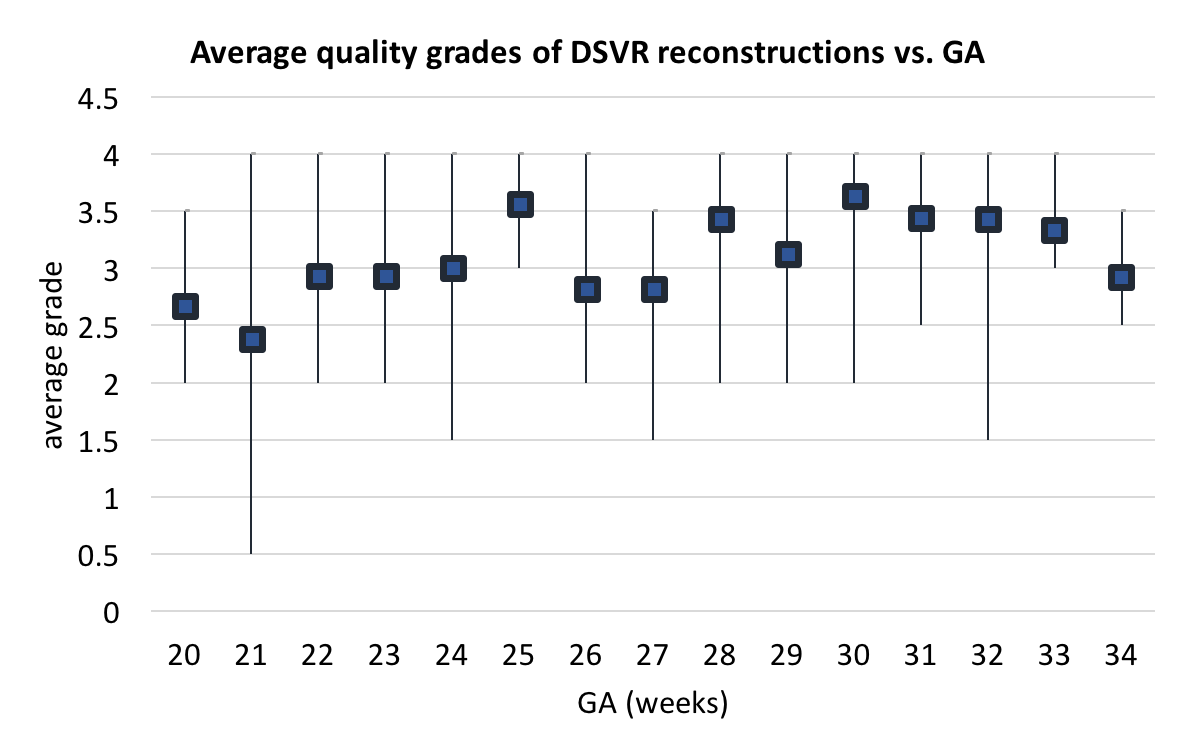}
\caption{Quality of DSVR reconstructions of fetal body region for 100 reconstructed iFIND cases vs. GA. Squares represent the average grade per week of GA, and bars represent the range of values.}
\label{fig_11}
\end{figure}

 To assess the applicability of DSVR across gestational ages, we performed qualitative evaluation for 100 iFIND fetal body cases randomly selected from 20-34 weeks GA range. The reconstructed volumes were graded by trained clinicians with respect to the image quality in [0; 4] range (4 corresponding to high quality). Volumes with grades $\geq2$ were considered to have sufficient quality for further analysis and interpretation. The distribution of grades for each week of GA is presented in Fig.~\ref{fig_11}. Average grades per week of GA varied within 2.5 - 3.5 range ($3.09 \pm 0.78 $). The primary causes of lower grades were motion for younger subjects and low SNR for older subjects. Only $ 6\% $ of all cases failed (grades $< 2$) due to severe motion which could not be resolved by gradient descent based FFD registration. Further details of the analysis are presented in Supplementary material.

\section{Discussion and Conclusions}

\par We proposed and implemented a novel DSVR method for compensation of non-rigid motion in fetal MRI. It allows reconstruction of high resolution 3D volumes from multiple stacks of slices affected by non-rigid motion. Therefore, DSVR extends application of slice-to-volume reconstruction to fetal body and placenta. 

\par Unlike the conventional rigid SVR methods, DSVR is capable of correction of local deformations of organs caused by bending and stretching. Correction of both in- and out-of-plane non-rigid motion is ensured by registration of the volume to slices rather than slices to volume. The challenge of the absence of a 'stable' shape is addressed by hierarchical FFD SVR scheme initialised by one of the stacks that gradually converges to a stable state. The fact that the stacks have different orientations helps to prevent overfitting to motion artefacts. In addition, structure-based outlier rejection step is introduced to minimise the impact of misregistration errors on the reconstructed volume. 

\par The method was quantitatively evaluated on 20 fetal MRI datasets from 28-31 weeks GA range. This age range was chosen due to high incidence of stretching and bending, whereas large rotation and translation motion is less prevalent compared to younger subjects \cite{Nowlan2015}. Comparison to the state-of-the-art solutions showed that DSVR surpasses both SVR and PVR methods for minor and severe motion datasets. This was further confirmed by an additional experiment with simulated non-rigid motion. DSVR reconstructed 3D fetal body volumes are characterised by well defined features and texture of the spine, heart and abdominal organs. 
\par The current implementation of DSVR is not designed for correction of large amplitude motion, including, large rotations and bending, due to the limited capture range of the employed gradient descent based optimisation methods, and therefore relies on outlier rejection in such cases. The qualitative study of DSVR reconstruction of fetal body for 100 iFIND cases from 20-34 weeks GA range showed that large amplitude motion primarily affects datasets of subjects under $25$ weeks GA. In future, this limitation can be addressed by application of CNN-based methods, as already proposed for rigid SVR fetal brain reconstruction \cite{Hou2018, Salehi2019}.

\par We further demonstrated that DSVR outperforms both SVR and PVR for 3D placenta reconstruction. In future, introducing additional decoupling of maternal motion would potentially improve correction of large amplitude motion. 

\par Although DSVR reconstructed volumes can be used for qualitative analysis, the question of volume-preservation for DSVR reconstruction still remains open. Quantitative measurements performed on DSVR reconstructed volumes can be influenced by various factors such as the position and shape of the fetal body in the template, number of stacks or CP spacing. This limitation should be addressed in future by introducing further model-based constrains on deformation fields along with automation of template selection and masking steps.

\appendices
\section{Supplementary Data}

Supplementary material presents the details of qualitative analysis of DSVR reconstruction quality for 100 iFIND cases, the full versions of Fig.~\ref{fig_5},~\ref{fig_6},~\ref{fig_8}, DSVR reconstruction of the whole uterus for twin pregnancy case, and visual representation of the DSVR algorithm.

\section*{Acknowledgment}

\par Thank you to Matthew Fox, Joanna Allsop, Ana Gomes and Elaine Green for their oversight during the scanning of volunteers and patients. Thank you to Jacqueline Matthew, Milou van Poppel and Johannes Steinweg for grading DSVR reconstruction quality for the investigated iFIND cases.

\par The iFIND project data used in this research were collected subject to the informed consent of the participants. The MIRTK library was used under the Apache License, Version 2.0. The original IRTK reconstruction code was used under the creative commons public license from IXICO Ltd. The views expressed are those of the authors and not necessarily those of the NHS, the NIHR or the Department of Health.

\bibliographystyle{IEEEtran}
\bibliography{main-references.bib}

\end{document}